\documentstyle[11pt,epsf]{article}

\newcommand{\be}{\begin{equation}}
\newcommand{\ee}{\end{equation}}
\newcommand{\nn}{\nonumber}
\newcommand{\beba}{\begin{equation}\begin{array}{lcl}}
\newcommand{\eaee}{\end{array}\end{equation}}
\newcommand{\bea}{\begin{eqnarray}}
\newcommand{\eea}{\end{eqnarray}}
\newcommand{\ba}{\begin{array}}
\newcommand{\ea}{\end{array}}

\newcommand{\ns}{\normalsize}
\newcommand{\refs}[1]{(\ref{#1})}

\def\abar{{\bar{a}}}
\def\bbar{{\bar{b}}}
\def\cbar{{\bar{c}}}

\def\tr{{\rm tr}}

\def\agut{\alpha_{\rm GUT}}

\newcommand{\orbav}[1]{\big<{#1}\big>_{11}}

\def\a{\alpha}
\def\b{\beta}

\def\c{\chi}
\def\d{\delta}
\def\e{\epsilon}

\def\f{\phi}

\def\k{\kappa}
\def\l{\lambda}
\def\m{\mu}
\def\n{\nu}
\def\o{\omega}
\def\p{\pi}

\def\r{\rho}
\def\s{\sigma}

\def\G{\Gamma}
\def\J{\Psi}
\def\L{\Lambda}
\def\O{\Omega}

\begin{document}


\begin{titlepage}
\vspace{-2cm}
\title{\hfill{\ns UPR-785T\\}
       \hfill{\ns hep-th/9711197\\[.5cm]}
       {\Large\bf Gaugino Condensation in M--theory on $S^1/Z_2$}}
\author{Andr\'e
        Lukas$^1$\setcounter{footnote}{0}\thanks{Supported in part by Deutsche
        Forschungsgemeinschaft (DFG).}~~,
        Burt A.~Ovrut$^1\, $\setcounter{footnote}{3}\thanks{Supported
        in part by a Senior Alexander von Humboldt Award}~~and Daniel
         Waldram$^2$\\[0.5cm]
        {\ns $^1$Department of Physics, University of Pennsylvania} \\
        {\ns Philadelphia, PA 19104--6396, USA}\\[0.3cm]
        {\ns $^2$Department of Physics}\\
        {\ns Joseph Henry Laboratories, Princeton University}\\
        {\ns Princeton, NJ 08544, USA}}
\date{}
\maketitle

\begin{abstract}
In the low energy limit of  for M--theory on $S^1/Z_2$, we calculate the
gaugino condensate potential in four dimensions using the background solutions
due to Ho\v rava. We show that this potential is free of delta--function
singularities and has the same form as the potential in the weakly coupled
heterotic string. A general flux quantization rule for the three--form field of
M--theory on $S^1/Z_2$ is given and checked in certain limiting cases.
This rule is used to fix the free parameter in the potential originating
from a zero mode of the form field. Finally, we calculate soft supersymmetry
breaking terms. We find that corrections to the K\"ahler potential and the
gauge kinetic function, which can be large in the strongly coupled region,
contribute significantly to certain soft terms. In particular, for
supersymmetry breaking in the $T$--modulus direction, the small values
of gaugino masses and trilinear couplings that occur in the weakly coupled,
large radius regime are enhanced to order $m_{3/2}$ in M--theory. The
scalar soft masses remain small even, in the strong coupling M--theory limit. 
\end{abstract}

\thispagestyle{empty}
\end{titlepage}


\section{Introduction}

New knowledge about the strongly coupled limits of heterotic string
theory has opened new possibilities for building realistic low-energy
models. Ho\v rava and Witten~\cite{hw1,hw2} have suggested that the
strongly coupled $E_8\times E_8$ heterotic string is described by M-theory
on a $S^1/Z_2$ orbifold, where the low-energy effective theory is
eleven-dimensional supergravity together with one set of $E_8$ gauge 
fields on each of the two orbifold fixed
hyperplanes~\cite{hw1,hw2}. Further, compactifying on a Calabi-Yau 
threefold, Witten has shown that it is possible to produce an $N=1$
theory in four dimensions, where the tree level gravitational and
gauge couplings match their known values~\cite{w}. One finds that the
Newton constant and grand-unified coupling constant are given by
\be 
  G_N = \frac{\k^2}{16\pi^2V\rho} \qquad
  \agut = \frac{\left(4\pi\k^2\right)^{2/3}}{2V} \; , \label{couplings}
\ee
where $\k$ is the eleven-dimensional gravitational coupling constant,
$V$ is the volume of the Calabi-Yau space and $\pi\r$ is the length of the
orbifold interval. The GUT scale is set by the size of the Calabi-Yau
three--fold, so that one takes $V^{-1/6}$ to be about $10^{16}$ GeV while the
eleven-dimensional Planck scale is about a factor of two larger. The
orbifold scale, $(\pi\r)^{-1}$, is approximately a factor of ten
smaller~\cite{bd}. Thus, with increasing energy, the universe appears first
four-, then five- and finally eleven-dimensional. 

As always, one must address how the supersymmetry of the
four-dimensional effective theory is eventually broken. A standard
mechanism in weakly coupled string theory has been supersymmetry breaking via
gaugino condensation in a hidden sector~\cite{drsw}. That such a mechanism
could also work in the strongly coupled model was, first demonstrated
in the full eleven-dimensional theory by Ho\v rava~\cite{hor}. The
purpose of the present paper is to discuss this approach in the context of
deriving the effective four-dimensional theory. In particular, we
obtain the form of the four-dimensional gaugino potential directly
from the M--theory effective action. We address the r\^ole of flux
quantization in restricting the form of the gaugino potential. Finally,
we discuss the structure of the soft supersymmetry breaking
terms. These depend crucially on the form of the low-energy K\"ahler
potential and superpotential, which were calculated directly from
M--theory in a previous paper~\cite{low}. 

The question of gaugino condensation in the low-energy effective
action of M--theory on $S^1/Z_2$ has already been addressed in various
places in the literature~\cite{LLN1,DG,AQ,hp,lt,D,EFN}. However, we believe a
number of points require clarification in these discussions, and
in particular and most importantly, a clear derivation of the gaugino
potential from the M theory effective theory has thus far been
missing. 

It is important in this discussion to be clear about the scale at
which the condensate forms. There are two distinct cases, with the
scale either above or below the scale set by the size of the orbifold
interval. In the former case, the theory is still effectively
five-dimensional when the condensate forms. The corresponding
gaugino potentials, and the question of moduli stabilization, should
then be discussed in the context of a full five-dimensional
theory~\cite{inprep}. A very interesting property of this case is
that, as pointed out by Ho\v rava~\cite{hor}, locally the condensate
does not break supersymmetry. However, there is a global obstruction
to preserving supersymmetry everywhere across the orbifold
interval. Consequently, only once one drops below the orbifold
threshold to a truly four-dimensional theory is the supersymmetry
breaking seen locally. In this case, in contrast to the conventional
picture, the scale of low-energy supersymmetry breaking may then not
set by the condensate scale, but instead by the orbifold
threshold. Since this scenario requires a five-dimensional
description, we will not pursue it further here.

In the other case, the theory becomes four-dimensional before the
gauginos condense. In this case, the supersymmetry breaking scale is
set, as usual, by the condensate scale, and is independent of the
orbifold size, other than through the fact that the four-dimensional
Planck scale depends on $\r$. This is the scenario
the majority of discussions have considered, since it is the
phenomenologically reasonable one. (One exception is Antoniadis and
Quiros~\cite{AQ}, who argue that the orbifold threshold is actually
down at $10^{12}$ GeV, and take the former scenario. They claim the
effects of the condensate in the detailed five-dimensional theory are
reproduced in four dimensions by a Scherk-Schwarz compactification.) It
is also worth noting that, in this second case, there is no way of
resolving the separation of the two orbifold planes since all momenta
are below the orbifold threshold. Thus, although the standard model
fields may come from one plane and the hidden sector fields live on
the other, there is no sense in the four-dimensional theory that we
are ``living on one plane''. 

In deriving the gaugino potential in the second scenario, it is
crucial to observe, as was stressed in~\cite{low} and has been missed
in other discussions, that one cannot
make a conventional dimensional reduction. In general, it is not
possible to simply excite fields which are independent of the orbifold
coordinate. In the Ho\v rava-Witten description, the fields on the
orbifold planes act as sources for the eleven-dimensional supergravity
fields. Only when the sources on the two planes are matched can the
bulk fields be taken to be independent of $x^{11}$. As a result, one
is always led to consider truncations with non-trivial $x^{11}$
dependence. It is the orbifold averages of these bulk fields which
lead to non-trivial couplings in the low-energy effective
action. For instance, they are the source of the Chern-Simons terms in
four-dimensions~\cite{low}, and here they are the
source of the gaugino condensate potential. It might appear from the
form of the eleven-dimensional action that the potential arises from
delta-function terms localized at the orbifold planes. However, as we
will show, in fact, all terms in the eleven-dimensional action are
smooth and the potential arises through the average of an $x^{11}$
dependent background. 

Our essential result will be that the resulting gaugino potentials are
the same as in the weakly coupled limit. In a previous
paper~\cite{low}, we calculated the form of the rest of the effective
action. One finds that it again agrees in form with the
large Calabi-Yau limit of the one-loop weak calculation,
barring one term which, while probably present in the weakly coupled
limit, appears previously to have been ignored. One finds the K\"ahler
potential, superpotential and gauge kinetic functions are given in
terms of the usual $S$, $T$ and charged matter moduli $C$, by 
\bea
 K &=& -\ln (S+\bar{S}-\b|C|^2/2)-3\ln (T+\bar{T}-|C|^2) \nn \\
 W &=& k\, d_{pqr}C^pC^qC^r \nn \\
 f^{(1)} &=& S+\b T \label{KWf}\\
 f^{(2)} &=& S-\b T \nn \; ,
\eea
where $k^2=4\agut/G_{\rm N}$. In the strongly coupled limit, one finds
$\b=\frac{\sqrt{2}\p\r}{16}\a_0=O(\k^{2/3}\r/V^{2/3})$ where $\a_0$ is
defined as in integral over the Calabi--Yau space~\cite{w,low}. In the weak
limit, $\b$ characterizes the loop correction and is small, while in the
strongly coupled limit it can be of order one. The 
previously neglected term is the correction $\b|C|^2$ in the K\"ahler
potential. Thus, while the form of the gaugino potentials are the same as in
the weakly coupled limit, since $\b$ need not be small we will find that this
term can considerably alter the soft terms resulting from supersymmetry
breaking. That the corresponding terms in the gauge kinetic functions
can significantly raise the gaugino masses compared with the weak
limit was previously noted in~\cite{hp}.

\section{Review of M--theory on $S^1/Z_2$}

To set the stage for the discussion of gaugino condensation, we would like
to briefly review the low energy theory for M--theory on
$S^1/Z_2$~\cite{hw1,hw2}, and its solutions appropriate for a reduction
to four dimensions~\cite{w,hor}.

Our conventions are as follows. We use indices $I,J,K,\ldots = 0,\ldots ,9,11$
for the full 11--dimensional space $M_{11}$. The orbifold $S^1/Z_2$ with
radius $\r$ is parameterized by $x^{11}\in [-\p\r ,\p\r ]$ so that the
two 10--dimensional fixed hyperplanes $M_{10}^{(1)}$, $M_{10}^{(2)}$ are
located at $x^{11}=0$ and $x^{11}=\p\r$. Occasionally, we will find it useful
to work in the boundary picture where only one half of the orbifold $S^1/Z_2$
is considered and the hyperplanes $M_{10}^{(i)}$ are viewed as boundaries of
$M_{11}$. The coordinate $x^{11}$ is then restricted to $x^{11}\in [0,\p\r ]$.
The 10--dimensional space transverse to the orbifold is labeled by barred
indices $\bar{I},\bar{J},\bar{K},\ldots=0,\ldots,9$. For the reduction to
four dimensions, we introduce indices $A,B,C,\ldots=4,\ldots,9$ for the
Calabi--Yau space and four--dimensional indices $\mu,\nu\ldots=0,\ldots,3$.
Holomorphic and antiholomorphic indices on the Calabi--Yau space are
denoted by $a,b,c,\ldots$ and $\abar,\bbar,\cbar,\ldots$, respectively.

The effective action for M--theory on $S^1/Z_2$, up the order $\k^{2/3}$
in the 11--dimensional Newton constant $\k$, is given by~\cite{hw2}
\bea
 S &=& \frac{1}{2\k^2}\int_{M_{11}}\sqrt{-g}\left[ -R-\frac{1}{24}G_{IJKL}
       G^{IJKL}-\bar{\J}_I\G^{IJK}D_J\left(\frac{\O +\hat{\O}}{2}\right)\J_K
       \right.\nn \\
    && \qquad\qquad\qquad\qquad
       -\frac{\sqrt{2}}{192}(\bar{\J}_I\G^{IJKLMN}\J_N+12\bar{\J}^J\G^{KL}\J^M)
       (G_{JKLM}+\hat{G}_{JKLM})\nn \\
    && \left. \qquad\qquad\qquad\qquad
       -\frac{\sqrt{2}}{1728}\e^{I_1...I_{11}}C_{I_1I_2I_3}
       G_{I_4...I_7}G_{I_8...I_{11}}\right]\nn \\
    && -\frac{1}{8\p\k^2}\left(\frac{\k}{4\p}\right)^{2/3}\sum_{i=1,2}
       \int_{M_{10}^{(i)}}\sqrt{-g}\,\tr\,\left[F^{(i)}_{\bar{I}\bar{J}}
       F^{(i)\bar{I}\bar{J}}+2\bar{\c}^{(i)}\G^{\bar{I}}D_{\bar{I}}
       (\hat{\O})\c^{(i)}\right.\nn \\
    && \left.\qquad\qquad\qquad\qquad\qquad
       +\frac{1}{2}\bar{\J}_{\bar{I}}\G^{\bar{J}\bar{K}}
      \G^{\bar{I}}(F^{(i)}_{\bar{J}\bar{K}}+\hat{F}^{(i)}_{\bar{J}\bar{K}})
      \c^{(i)}-\frac{\sqrt{2}}{12}\bar{\c}^{(i)}\G^{\bar{I}\bar{J}\bar{K}}
      \c^{(i)}\hat{G}_{\bar{I}\bar{J}\bar{K}11}\right] \label{S}
\eea 
The bulk fields in this action are the 11--dimensional metric $g_{IJ}$,
the three--form $C_{IJK}$ with bulk field strength 
$G_{IJKL}=24\partial_{[I}C_{JKL]}$ and the gravitino $\J_I$. The two
$E_8$ gauge fields $A^{(i)}_{\bar{I}}$, $i=1,2$ with field strengths
$F^{(i)}_{\bar{I}\bar{J}}$ and their gaugino superpartners $\c^{(i)}$
live on the 10--dimensional hyperplanes $M_{10}^{(i)}$. The supercovariant
objects $\hat{\O}$, $\hat{F}_{\bar{I}\bar{J}}$ and $\hat{G}_{IJKL}$ are
defined in ref.~\cite{hw2}, but will not be needed explicitly in this paper.
The above action has to be supplemented with the Bianchi identity
\be
 (dG)_{11\bar{I}\bar{J}\bar{K}\bar{L}} = -\frac{1}{2\sqrt{2}\pi}
    \left(\frac{\k}{4\pi}\right)^{2/3} \left[
    J^{(1)} \d (x^{11})+ J^{(2)}\d (x^{11}-\pi\r )\right]
    _{\bar{I}\bar{J}\bar{K}\bar{L}} \; .\label{Bianchi}
\ee
for $G$, where the sources are defined by 
\be
 J^{(i)} = \left( \tr\,F^{(i)}\wedge F^{(i)} 
          - \frac{1}{2}\tr\,R\wedge R \right)=d\o^{(i)}\; . \nn \\
\ee
The three--forms $\o^{(i)}$ can be expressed in terms of the Yang--Mills
and Lorentz Chern--Simons forms $\o^{{\rm YM}(i)}$ and $\o^{\rm L}$ as
\be
  \o^{(i)} = \o^{{\rm YM}(i)} - \frac{1}{2}\o^{\rm L} \; ,
  \label{CS_def}
\ee
The bulk fields should respect the $Z_2$ orbifold symmetry,
which implies for the bosonic fields that $g_{\bar{I}\bar{J}}$, $g_{11,11}$,
$G_{\bar{I}\bar{J}11}$ are even and $g_{\bar{I}11}$,
$C_{\bar{I}\bar{J}\bar{K}}$ are odd.

For this theory, the existence of background solutions suitable for
compactification to four dimensions with preserved $N=1$ supersymmetry
has been demonstrated in ref.~\cite{w}. These solutions, based on the
standard embedding of the spin connection into one of the $E_8$ gauge groups,
are of the form $M_{11}=S^1/Z_2\times X\times M_4$ where $X$ is a 
deformed Calabi--Yau space and $M_4$ is four--dimensional Minkowski space.
The deformation of the Calabi--Yau space, as well as the existence of a
nonzero form $G=G^{(W)}$ in those solutions, is due to the fact that the
sources in the Bianchi identity~\refs{Bianchi} do not vanish, unlike in
the weakly coupled case where this is guaranteed by the standard embedding.
The modification of the solutions in the presence of gaugino
condensation has been worked out by Ho\v rava~\cite{hor}. Since it is for
those backgrounds that we will later compute the gaugino condensate potential,
we will briefly review the main results of this work.

\vspace{0.4cm}

A crucial observation is that, similar to the weakly coupled case, the terms
in the action which contain gaugino bilinears can, together with the $G^2$
term, be grouped into a perfect square
\be
 S^{(\c )} = -\frac{1}{48\k^2}\int_{M_{11}}\sqrt{-g}\tilde{G}_{IJKL}
             \tilde{G}^{IJKL}\label{ps}
\ee
with
\bea
 \tilde{G}_{\bar{I}\bar{J}\bar{K}\bar{L}}
     &=& G_{\bar{I}\bar{J}\bar{K}\bar{L}} \nn\\
 \tilde{G}_{\bar{I}\bar{J}\bar{K}11} &=& G_{\bar{I}\bar{J}\bar{K}11}
   -\frac{\sqrt{2}}{16\p}\left(\frac{\k}{4\p}\right)^{2/3}\left(
   \d (x^{11})\o^{(\c ,1)}+\d (x^{11}-\p\r ) \o^{(\c ,2)}
   \right)_{\bar{I}\bar{J}\bar{K}} \label{Gtdef}
\eea
and the gaugino bilinears
\be
 \o^{(\c ,i)}_{\bar{I}\bar{J}\bar{K}}=\tr\,\bar{\c}^{(i)}
             \G_{\bar{I}\bar{J}\bar{K}}\c^{(i)}\; .
\ee
The terms proportional to $(\o^{(\c ,i)})^2$ arising from this square are of
order $\k^{4/3}$ and have, therefore, been omitted from the
action~\refs{S}. In ref.~\cite{hw2} they were shown to follow from the
cancellation of the supersymmetry variation at order $\k^{4/3}$, with
exactly the right coefficient to fit into the perfect square. Note, however,
that these terms are proportional to $(\d (x^{11}))^2$ and are, therefore, not
quite well--defined. In ref.~\cite{hw2}, it was argued that this problem is
due to the finite thickness of the boundary, which has not been taken into
account in constructing the effective theory. 

In terms of the redefined form field $\tilde{G}$, the supersymmetry
transformation of the gravitino reads
\bea
 \d\J_{\bar{I}} &=& D_{\bar{I}}\eta +\frac{\sqrt{2}}{288}\tilde{G}_{JKLM}
                    ({\G_{\bar{I}}}^{JKLM}-8\d_{\bar{I}}^J\G^{KLM})\eta
                    \label{k10}\\
 \d\J_{11} &=& D_{11}\eta +\frac{\sqrt{2}}{288}\tilde{G}_{JKLM}
                    ({\G_{11}}^{JKLM}-8\d_{11}^J\G^{KLM})\eta\nn \\
           && +\frac{1}{192\p}\left(\frac{\k}{4\p}\right)^{2/3} 
              \left[\d (x^{11})\o^{(\c ,1)}+\d (x^{11}-\p\r )
              \o^{(\c ,2)}\right]_{\bar{I}\bar{J}\bar{K}}\G^{\bar{I}
              \bar{J}\bar{K}}\eta
              \label{k11}
\eea
where the spinor $\eta$ is subject to the $Z_2$ restriction
$\eta (-x^{11})=\G^{11}\eta (x^{11})$.
The equation of motion for $\tilde{G}$ to be derived from eq.~\refs{ps}
\footnote{We do not write the contribution from the Chern--Simons term
in~\refs{S} to the equation of motion since it vanishes for the
solutions we will consider.}, and the modified Bianchi identity~\refs{Bianchi},
are given by
\bea
 D_I\tilde{G}^{IJKL} &=& 0 \label{Geom}\\
 (d\tilde{G})_{11\bar{I}\bar{J}\bar{K}\bar{L}} &=& -\frac{1}{2\sqrt{2}\p}
   \left(\frac{\k}{4\p}\right)^{2/3}\left[ J^{(1)} \d (x^{11})+ 
   J^{(2)}\d (x^{11}-\pi\r )\right]_{\bar{I}\bar{J}\bar{K}\bar{L}} \nn \\
  &&-\frac{\sqrt{2}}{16\p}\left(\frac{\k}{4\p}\right)^{2/3}\left[
  J^{(\c ,1)}\d (x^{11})+J^{(\c ,2)}\d (x^{11}-\p\r )\right]_{\bar{I}\bar{J}
  \bar{K}\bar{L}} \label{Btilde}
\eea
where the sources $J^{(\c ,i)}$ induced by the gaugino condensates are defined
as
\be
 J^{(\c ,i)} = d\o^{(\c ,i)}\; .
\ee
If these condensates are covariantly constant, that is $J^{(\c ,i)}=0$,
the additional source terms in the above Bianchi identity disappear and
$\tilde{G}$ is governed by the same equations as $G$ satisfied with vanishing
gaugino condensates. In this case, which we consider from now on, one can
find a solution of the theory
if one uses Witten's result for the metric~\cite{w} and if one sets
$\tilde{G}$ equal to the form field $G^{(W)}$ of Witten's solution. Clearly,
if the condensates are switched off in the solution constructed in this way,
one goes back to a supersymmetry preserving configuration which satisfies
the Killing spinor equations $\d\J_I=0$ for a certain spinor
$\tilde{\eta}$. For nonzero condensates, of course, one cannot, a priori,
expect to be able to fulfill those equations since supersymmetry might be
broken. However, since the 10--dimensional part of the gravitino
variation~\refs{k10} is unchanged by the condensate, the condition
$\d\J_{\bar{I}}=0$ can be fulfilled by the original spinor $\tilde{\eta}$.
One can try to satisfy the remaining equation $\d\J_{11}=0$ by modifying
the spinor $\tilde{\eta}$ to $\eta '$ (such that $\d\J_{\bar{I}}=0$
remains true). From eq.~\refs{k11}, this leads to the condition
\be
 \partial_{11}(\eta '-\tilde{\eta}) = -\frac{1}{192\p}\left(\frac{\k}{4\p}
    \right)^{2/3}\left[ \o^{(\c ,1)}\d (x^{11})+\o^{(\c ,2)}\d(x^{11}-\p\r )
    \right]_{\bar{I}\bar{J}\bar{K}}\G^{\bar{I}\bar{J}\bar{K}}\eta_0
 \label{kd}
\ee
where $\eta_0$ is the Killing spinor of the original Calabi--Yau space. As
pointed out by Ho\v rava, this equation has a local solution for $\eta '$
everywhere but, in general, there exists no global solution. To see this
more explicitly, we find it useful to rewrite eq.~\refs{kd} in the
boundary picture. In this picture, the delta--function sources on the fixed
hyperplanes turn into boundary conditions and we have
\bea
 \partial (\eta '-\tilde{\eta}) &=& 0 \nn \\
 \left. (\eta '-\tilde{\eta})\right|_{x^{11}=0} &=& -\frac{1}{384\p}
        \left(\frac{\k}{4\p}\right)^{2/3}\o^{(\c ,1)}_{\bar{I}\bar{J}\bar{K}}
        \G^{\bar{I}\bar{J}\bar{K}}\eta_0 \\
 \left. (\eta '-\tilde{\eta})\right|_{x^{11}=\p\r} &=& \frac{1}{384\p}
        \left(\frac{\k}{4\p}\right)^{2/3}\o^{(\c ,2)}_{\bar{I}\bar{J}\bar{K}}
        \G^{\bar{I}\bar{J}\bar{K}}\eta_0\nn\; .
\eea
Obviously, $\eta '-\tilde{\eta} =\mbox{const}$ is a solution everywhere 
locally but a problem arises if one tries to match it to both boundary
values. Clearly, this requires
\be
 (\o^{(\c ,1)}+\o^{(\c ,1)})_{\bar{I}\bar{J}\bar{K}}\G^{\bar{I}\bar{J}\bar{K}}
 \eta_0 = 0 \label{unbroken}
\ee
which is fulfilled for equal but opposite condensates. In general,
eq.~\refs{unbroken} is not satisfied and supersymmetry is broken.
This breaking mechanism can be called global in the sense that information
from both hyperplanes is required in order to realize the breaking.

\vspace{0.4cm}

Several questions arise from the above discussion. Clearly, one would like
to know the low energy potential that describes the dynamics of this
type of supersymmetry breaking, and one would like to understand how the global
nature of this breaking is encoded in the potential. From what was said so
far, however, the computation of this potential leads to a number of problems.
As already mentioned, the $(\o^{(\c ,i)})^2$ term from the perfect
square~\refs{ps}, which gives rise to a potential in the analogous
weakly coupled setting, appears with a factor $(\d (x^{11}))^2$. If the low
energy potential would indeed directly arise from this term, it would be
proportional to $\d (0)$ and therefore be poorly defined. Indeed, this has
been found in ref.~\cite{lt}, where the perfect square~\refs{ps} has simple
been multiplied out to compute the low energy potential. In essence, such a
result implies that the limitations of the effective 11--dimensional theory
do not allow for a proper calculation of the low energy gaugino potential.
On the other hand, following Ho\v rava, we have set the modified form
$\tilde{G}$ equal to Witten's background $G^{(W)}$, which is smooth.
In this approach, the reduction of the perfect square~\refs{ps} does not
produce any $\d (0)$ singularities in the low energy action. Curiously,
however, the gaugino condensate has also disappeared from the action and
it would seem that no potential is generated at all.

In the following, we are going to resolve these and other problems and we will
show that a well--defined gaugino condensate potential, which encodes the
characteristics of the global breaking mechanism, can be computed by reduction
of the 11--dimensional action.

\section{Computation of the potential}

The gaugino condensate potential which we are going to compute in this
section will be part of a low energy theory otherwise specified by the
K\"ahler potential, the superpotential and the gauge kinetic functions given
in eq.~\refs{KWf}. These quantities were systematically derived in
ref.~\cite{low} by reducing the effective theory~\refs{S} on Witten's
deformed Calabi--Yau background to four dimensions. Let us first review
the essential steps of this derivation so that we can, later on, precisely
specify the modifications needed to incorporate gaugino condensation.

The bulk field configurations used for the reduction can be schematically
written as
\bea
 g_{IJ} &=& g_{IJ}^{(0)}+g_{IJ}^{(1)}+g_{IJ}^{(B)} \label{g}\\
 G_{IJKL} &=& G^{(0)}_{IJKL}+G^{(1)}_{IJKL}+G^{(B)}_{IJKL}+G^{(W)}_{IJKL}
 \label{G}
\eea
where the contributions to zeroth order in $\k$ are explicitly given by
\bea
 g^{(0)}_{\m\n} = \bar{g}_{\m\n}\; ,&g^{(0)}_{AB}=e^{2a}\O_{AB}\; ,&
 g^{(0)}_{11,11} = e^{2c} \label{g0}\\
 G^{(0)}_{\m\n\r 11} = 3\partial_{[\m} B_{\n\r ]}\; ,
 &G^{(0)}_{\m AB 11} = \partial_\m\f\, \o_{AB}&\; .
\eea
In these expressions, $\O_{AB}$ and $\o_{AB}$ are the metric and the K\"ahler
form of the (undeformed) Calabi--Yau space with volume $V$. The generic moduli
$a$ and $c$ measure the radii of the Calabi--Yau space and the orbifold,
respectively, and the two--form $B_{\m\n}$ and the scalar $\f$ are their
bosonic superpartners. The Einstein frame metric $g_{\m\n}$ is related to
$\bar{g}_{\m\n}$ by the Weyl rotation
\be
 \bar{g}_{\m\n} = e^{-6a-c}g_{\m\n}\; .\label{Weyl}
\ee
With the generic gauge matter field $C^p$ in the
fundamental representation ${\bf 27}$ of $E_6$ defined by
\be
 A^{(1)}_b=\bar{A}_b+{w_b}^cT_{cp}C^p\; . \label{gauge_matter}
\ee
(here $\bar{A}_b$ is the embedded spin connection), we can express the
low energy moduli fields $S$ and $T$ as
\be
 S=e^{6a}+i\sqrt{2}\s+\frac{1}{2}\b|C|^2\; ,\quad
  T = e^{c+2a}+i\sqrt{2}\f+\frac{1}{2}|C|^2\; ,
\ee
where the dual $\s$ of the field strength
$H_{\m\n\r}=3\partial_{[\m} B_{\n\r]}$ is defined by
$H_{\m\n\r} = e^{-12a}{\e_{\m\n\r}}^\s\partial_\s\s$. The constant $\b$
is the same one that appears in the K\"ahler potential and the gauge kinetic 
functions~\refs{KWf}, and its value can be explicitly computed for a given
Calabi--Yau manifold~\cite{w,low}. The bosonic field content of the low
energy theory is completed by the observable gauge fields $A^{(1)}_\m$
which, due to the standard embedding, belong to the gauge group $E_6$, and
the hidden sector gauge fields $A^{(2)}_\m$ with gauge group $E_8$.

The other contributions to the metric and the four--form in the
eqs.~\refs{g}, \refs{G} deserve an explanation. The metric correction
$g^{(1)}_{IJ}$ contains the moduli dependent order $\k^{2/3}$ correction 
to the Calabi--Yau background. Correspondingly, $G^{(1)}_{IJKL}$
represents the correction of the form zero modes due to this background
distortion. As already mentioned, $G^{(W)}$ is the form part of Witten's
solution needed to account for the sources in the Bianchi identity which are
generated by internal gauge fields. Corresponding terms $G^{(B)}_{IJKL}$
arise for source terms switched on by the external gauge fields. These
gauge fields also generate a nonvanishing stress energy on the boundaries,
which makes the compensating part $g^{(B)}_{IJ}$ of the metric necessary.
All these quantities, and their dependence on the generic moduli, have been
explicitly determined in ref.~\cite{low}. They have been proven to be
essential for a correct derivation of the low energy action.

\vspace{0.4cm}

Let us now incorporate gaugino condensates into the above scheme. In
analogy with the procedure for the background solution explained in
the previous section, one might try to set the full solution~\refs{G} for
the form equal to $\tilde{G}$. This is, however, not quite correct since
a condensate which is covariantly constant as part of the pure
background (and we will assume our condensates to be of that type) does
not remain covariantly constant once the moduli are promoted to fields
with explicit dependence on the four--dimensional coordinates. As a result,
though the equation of motion for $\tilde{G}$, eq.~\refs{Geom}, still
formally coincides with the equation of motion for $G$ in the absence of
a condensate, the Bianchi identity~\refs{Btilde} for $\tilde{G}$
receives  new sources from the condensate. We can account for these additional
sources by adding yet another piece, $G^{(\c )}_{IJKL}$, to the right hand
side of eq.~\refs{G}, so that $\tilde{G}$ takes the form
\be
 \tilde{G}_{IJKL} = G^{(0)}_{IJKL}+G^{(1)}_{IJKL}+G^{(B)}_{IJKL}+
                      G^{(W)}_{IJKL}+G^{(\c )}_{IJKL}
 \label{Gtilde}
\ee
Since the first four terms in this expression take care of all sources
in the Bianchi identity except the ones arising from condensates,
$G^{(\c )}_{IJKL}$ is subject to the following equations
\bea
 D_IG^{(\c )IJKL} &=& 0 \label{Gceom}\\
 (dG^{(\c )})_{11\bar{I}\bar{J}\bar{K}\bar{L}} &=& -\frac{\sqrt{2}}{16\p}
 \left(\frac{\k}{4\p}\right)^{2/3}\left[J^{(\c ,1)}\d (x^{11})+
 J^{(\c ,2)}\d (x^{11}-\p\r )\right]_{\bar{I}\bar{J}\bar{K}\bar{L}}
 \label{BGc}\; .
\eea
As has been pointed out in ref.~\cite{hw2}, a solution to these equations
should be free of delta--function singularities. Correspondingly, the first
four parts of $\tilde{G}$ do not contain any delta--function contributions
either, so that $\tilde{G}$ is a smooth object (in the
sense that it contains at most step functions) as already
pointed out in~\cite{hor}. Clearly, the solution can be expressed in terms
of the original form field $G$ via eq.~\refs{Gtdef} as well, but $G$
now contains a compensating delta--function contribution so as to make
$\tilde{G}$ smooth. The crucial point is that the action~\refs{ps}
can be entirely written in terms of the smooth object $\tilde{G}$ and,
therefore, leads to a perfectly well defined finite low energy effective
action without any $(\d (x^{11}))^2$ terms, no matter what specific
combination of $G$ and the condensates is used to describe the solution.
What we have really done is to define a new independent form field
$\tilde{G}$ such that the $(\d (x^{11}))^2$ terms are eliminates from the
action. The form field--gaugino coupling is then encoded in the new
Bianchi identity~\refs{Btilde}, in a way analogous to the Chern--Simons
terms. Since, as we have just argued, the solution for $\tilde{G}$ should
be smooth, the $(\d (x^{11}))^2$ singularity at order $\k^{4/3}$ is completely
removed.

To find an explicit solution for $G^{(\c )}$, we rewrite the eqs.~\refs{Gceom},
\refs{BGc} in the boundary picture as
\bea
 D_IG^{(\c )IJKL} &=& 0\nn \\
 dG &=& 0\nn \\
 \left. G_{\bar{I}\bar{J}\bar{K}\bar{L}}\right|_{x^{11}=0} &=&
   -\frac{\sqrt{2}}{32\p}\left(\frac{\k}{4\p}\right)^{2/3}
   J^{(\c ,1)}_{\bar{I}\bar{J}\bar{K}\bar{L}} \\
 \left. G_{\bar{I}\bar{J}\bar{K}\bar{L}}\right|_{x^{11}=\p\r} &=&
   \frac{\sqrt{2}}{32\p}\left(\frac{\k}{4\p}\right)^{2/3}
   J^{(\c ,2)}_{\bar{I}\bar{J}\bar{K}\bar{L}}\; .
\eea
These equations can be solved approximately to lowest order term in
a momentum expansion~\cite{low}, where term of the form $\r\partial J^{(i)}$
are neglected. This is justified~\footnote{This argument does not apply
to terms with derivatives in the Calabi--Yau direction since the Calabi--Yau
radius can be smaller than the orbifold radius. For covariantly constant
condensates, however, those terms vanish.} in the present case,
since the sources are provided by low energy fields with momenta far smaller
than the inverse orbifold radius $\r^{-1}$. The solution then reads
\bea
 G^{(\c )}_{\bar{I}\bar{J}\bar{K}11} &=& -\frac{\sqrt{2}}{32\p^2\r}
           \left(\frac{\k}{4\p}\right)^{2/3}\left[\o^{(\c ,1)}+\o^{(\c ,2)}
           \right]_{\bar{I}\bar{J}\bar{K}}+G'_{\bar{I}\bar{J}\bar{K}11} \\
 G^{(\c )}_{\bar{I}\bar{J}\bar{K}\bar{L}} &=& -\frac{\sqrt{2}}{32\p}
           \left(\frac{\k}{4\p}\right)^{2/3}\left[ J^{(\c ,1)}
           -\frac{x^{11}}{\p\r}(J^{(\c ,1)}+J^{(\c ,2)})
           \right]_{\bar{I}\bar{J}\bar{K}\bar{L}}+
           G'_{\bar{I}\bar{J}\bar{K}\bar{L}}
\eea
where $G'$ is an arbitrary zero mode; that is, a solution of the homogeneous
equations
\be
 D_I{G'}^{IJKL} = 0\; ,\qquad dG' = 0\; .\label{Gp}
\ee
In the terminology of ref.~\cite{gsw}, a topologically nontrivial $G'$
constitutes a ``nonzero'' zero mode which breaks supersymmetry explicitly.
Such a piece should be generally admitted in a background with
broken supersymmetry. It will, however, introduce some arbitrariness into
the gaugino condensate potential, which we are going to fix later on
by a flux quantization argument.

\vspace{0.4cm}

Let us now be more specific about the form of the condensate we are
going to consider. For the time being, we will allow for condensates on
both hyperplanes, though later on we will restrict the discussion to
the ``physical'' case of hidden $E_8$ condensates only. We use the
standard expression for covariantly constant condensates
\be
 \o^{(\c ,i)}_{abc} = \tr\, \bar{\c}\G_{abc}\c = \L^{(i)}\e_{abc}
 \label{cond} 
\ee
where the $\L^{(i)}$ represent the third powers of the condensation scales
(to be explicitly determined later) and are, therefore, viewed as functions
of the moduli $S$ and $T$. The nonvanishing components of the currents
$J^{(\c ,i)}$ then read
\be
 J^{(\c ,i)}_{\m abc} = \partial_\m\L^{(i)}\e_{abc}\; 
\ee
By means of eq.~\refs{Gp}, $G'$ is a harmonic four--form on
$S^1/Z_2\times X$. There exist $h^{2,1}+1$ independent such forms and
a basis is provided by
$\{ \e_{abc}, \o^{(i)}_{ab\bar{c}},i=1,\ldots ,h^{2,1}\}$
where $\o^{(i)}_{ab\bar{c}}$ are the $h^{2,1}$ harmonic $(2,1)$--forms of the
Calabi--Yau space $X$. Given the index structure of the condensate, only
the first of those forms can mix with the gaugino condensate in the low
energy potential and is, therefore, of interest to us. Consequently, we set
\be
 G'_{abc11} = -\frac{\sqrt{2}}{32\p^2\r}\left(\frac{\k}{4\p}\right)^{2/3}
              \l\e_{abc}
\ee
(with all other components vanishing) where $\l$ is a free
$x^{11}$--independent parameter. As a result, we find for $G^{(\c )}$
\bea
 G^{(\c )}_{abc11} &=& -\frac{\sqrt{2}}{32\p^2\r}\left(\frac{\k}{4\p}
                   \right)^{2/3}\left[ \L^{(1)}+\L^{(2)}+\l\right]\e_{abc}\; .
 \label{Gc} \\
 G^{(\c )}_{\m abc} &=& -\frac{\sqrt{2}}{32\p}\left(\frac{\k}{4\p}
                        \right)^{2/3}\partial_\m\left[\L^{(1)}-\frac{x^{11}}
                        {\p\r}(\L^{(1)}+\L^{(2)})\right]\e_{abc}
 \label{Gc1}
\eea
With these expressions, the field $\tilde{G}$ in eq.~\refs{Gtilde} is
completely determined in the presence of the condensates and can be
inserted into eq.~\refs{ps} to obtain the low energy potential. Clearly, one
contribution to the potential arises from the square of $G^{(\c )}_{abc11}$.
In addition, however, there can be mixing terms between $G^{(\c )}_{abc11}$
and the other parts of $\tilde{G}$ in eq.~\refs{Gtilde} which produce
potential terms. Inspection of those other parts as given in ref.~\cite{low},
shows that there is only one component with the correct index structure;
namely
\be
 G^{(B)}_{abc11} = -\frac{i}{\sqrt{2}\p^2\r}\left(\frac{\k}{4\p}\right)^{2/3}
                   d_{pqr}C^pC^qC^r\e_{abc}\; .
 \label{GB}
\ee
The corresponding mixing term is responsible for a cross term between
the matter fields and the gaugino condensate that should appear in the
scalar potential. For the discussion
of gaugino condensation, this term is usually neglected since $C$, as a matter
field, is thought of as fluctuating on small scales. It is, however, possible
that $C$ acquires a large vacuum expectation value $<C>$ which breaks the $E_6$
GUT symmetry. Though such an effect would be sizeable we will, in the
following, stick to the case $<C>=0$, for simplicity. The generalization
to include a nonzero $<C>$ is straightforward and does not change any of
the essential conclusions. With these remarks in mind, we use the
eqs.~\refs{ps}, \refs{g0}, \refs{Weyl} and \refs{Gc} to compute the
potential
\be
 S^{(\c )} = -\frac{V\p\r}{\k^2}\frac{1}{512\p^4\r^2}\left(\frac{\k}{4\p}
           \right)^{4/3}\int_{M_4}\sqrt{-g}e^{-12a-3c}|\L^{(1)}+
           \L^{(2)}+\l |^2\; .\label{pot}
\ee
We see that the potential can be expressed in terms of the condensates
$\L^{(i)}$ (whose functional dependence on the moduli will be inserted later)
and the arbitrary constant $\l$ which originates from the explicitly
supersymmetry breaking zero mode field $G'$. The condensate terms in
this potential do not arise directly from the corresponding terms in
11 dimensions, but rather as a part of the form field which is necessary to
interpolate between condensate source terms on the boundaries, as explained
in the introduction.
Therefore, the potential does not contain any delta--function singularities.
In the next section, we will argue that the free parameter $\l$ can be
fixed by a flux quantization argument in M--theory on $S^1/Z_2$, in
analogy with the weakly coupled case~\cite{rw}.

Before we come to this, let us discuss the relation of the above potential to
Ho\v rava's global breaking mechanism. Clearly, since we are working
in four dimensions, the global nature of the breaking (which is global
in the now invisible orbifold direction) cannot be really reflected in the
potential. To see the global form one would need to formulate a 
five-dimensional effective theory, where the dependence on the orbifold
coordinate is not integrated out. Despite this, there
is one characteristic of the global breaking which should persist in four
dimensions, namely the vanishing of supersymmetry breaking for equal
but opposite condensates. In view of the eqs.~\refs{unbroken} and
\refs{cond} such a situation, namely the existence of a global Killing spinor,
occurs for
\be
 \L^{(1)}+\L^{(2)} = 0\; . \label{unbroken1}
\ee
On the other hand, this condition is identical with the vanishing of
the gaugino part of the potential~\refs{pot} which proves the required
consistency. It is interesting to note that we could not arrive at such
a consistency by simply multiplying out the perfect square~\refs{ps} as has
been done in ref.~\cite{lt}. Since the cross term, being proportional
to $\d (x^{11})\d (x^{11}-\p\r )$, would (at least naively) vanish,
such a procedure leads to a potential proportional to
$(\L^{(1)})^2+(\L^{(2)})^2$. This is not necessarily zero if the
condition~\refs{unbroken1} is fulfilled.

\section{Flux quantization in M--theory on $S^1/Z_2$}

To proceed, it is necessary to establish the flux quantization condition for
Ho\v rava-Witten theory. The appropriate quantization condition in M-theory was
presented in ref.~\cite{wf}. It states that, for any closed four-cycle $C_{4}$
\begin{equation}
\zeta \int_{C_4}{ \frac{G}{2\pi}} + \int_{C_4}{ \frac{\lambda}{2}} = n
\label{eq:A}
\end{equation}
where $n$ is any integer, $G=dC$ locally,
$\lambda=\frac{1}{16\pi^{2}} R \wedge R$ and
$\zeta=\frac{1}{\sqrt{2}}(\frac{4\pi}{\kappa})^{\frac{2}{3}}$. As discussed in
that paper, this result applies to any closed four-cycle in Ho\v rava--Witten
theory that is homologous to a four-cycle on a single boundary plane.
Thus, closed four-cycles purely in the 11-dimensional bulk, or lying partially
or wholly in one boundary plane, satisfy (\ref{eq:A}). However, it is important
to note that there is another class of cycles in Ho\v rava--Witten theory;
namely, those that wind around the orbifold or, in the boundary picture, stretch
between both boundary planes. These new cycles are not closed in the boundary
picture.  Unlike closed cycles that, by definition, possess no boundary, 
these new cycles have a non-empty boundary three-cycle which is 
shared between both planes,
$\partial C_{4}=\partial C_{4}^{(1)}+\partial C_{4}^{(2)}$. Such cycles are not
homologous to closed four-cycles in the bulk or on one boundary and, hence, do
not necessarily satisfy equation (\ref{eq:A}). It is precisely the flux
quantization condition over these new cycles that is required 
in this paper to discuss the low energy effective action.\\

When one allows the new open 
four-cycles touching both boundary planes, in addition 
to the closed four-cycles in the bulk or touching a single boundary, we find
that Witten's flux quantization condition generalizes to
\begin{equation}
\zeta \int_{C_4}{ \frac{G}{2\pi}} + \int_{C_4}{ \frac{\lambda}{2}} 
+\frac{1}{16 \pi^{2}} \sum_{i=1}^{2} \int_{\partial
C_{4}^{(i)}}{\omega^{{\rm YM}(i)}} = n
\label{eq:B}
\end{equation}
where $\omega^{YM(i)}$ is the Yang-Mills Chern-Simons three-form on the $i$-th
boundary. In this paper, we will justify this flux quantization formula by
demonstrating that it is consistent with two limiting cases. First note, that
for a closed four-cycle either in the bulk or touching one of the boundary
planes, the Yang-Mills Chern-Simons term vanishes and we recover Witten's
expression (\ref{eq:A}). A second, far more non-trivial check is to consider a
Ho\v rava--Witten cycle that touches both boundaries. Furthermore, let us
choose the radius of the boundary cycles, $\partial C_{4}^{(i)}$, to be much
larger than the radius of the $S^{1}/Z_{2}$ orbifold. In this limit, it was
shown in ref.~\cite{low} that, at least locally, in the
ten-dimensional space, 
\begin{equation}
 G_{\bar I\bar J\bar K 11} \equiv H_{\bar I\bar J\bar K} =
3\partial_{[\bar I}B_{\bar J\bar K ]}
- \frac{1}{4\sqrt{2}\pi^2\rho}\left(\frac{\kappa}{4\pi}\right)^{2/3}
\left\{\omega^{{\rm YM}(1)}+\omega^{{\rm YM}(2)}-\omega^{\rm L}
\right\}_{\bar I\bar J\bar K}+ \Delta H_{\bar I\bar J\bar K}
\label{eq:C}
\end{equation}
where $\omega^{\rm L}$ is the Lorentz Chern-Simons three-form. 
The $B_{\bar J\bar K}$ and Chern-Simons forms are independent of
$x^{11}$. The $\Delta H_{\bar I\bar J\bar K}$ term is due to massive Kaluza-Klein
modes on the orbifold and is explicitly $x^{11}$ dependent. It is not
necessary to know its exact form, but only to recall that it must satisfy the
relation
\begin{equation}
  \orbav{\Delta H_{\bar I\bar J\bar K}} = 0
\label{eq:D}
\end{equation}
where $\orbav{\ldots}$ indicates the average over the 11-direction.
In this limit, we can evaluate the flux quantization condition. First, using
the fact that
\begin{equation}
\left(\tr R\wedge R\right)_{11\bar{I}\bar{J}\bar{K}} =  
    \left(d\o^{\rm L}\right)_{11\bar I \bar J \bar K}
\label{eq:E}
\end{equation}
equation (\ref{eq:B}) becomes
\begin{equation}
\zeta \int_{C_4}{ \frac{G}{2\pi}}  
+\frac{1}{16 \pi^{2}} \sum_{i=1}^{2} \int_{\partial
C_{4}^{(i)}}{(\omega^{{\rm YM}(i)}-\frac{1}{2}\omega^{{\rm L}(i)})} = n
\label{eq:F}
\end{equation}
Now, inserting expression (\ref{eq:C}) into this equation and using relation
(\ref{eq:D}), we find that the Chern-Simons terms exactly cancel and we are left
with the relation
\begin{equation}
\zeta \int_{C_4}{H^{0}}= 2\pi n
\label{eq:G}
\end{equation}
where locally $H^{0}=dB$. Now this limit, where the radius of the boundary
cycles is much larger than the orbifold radius, corresponds to the strong
coupling limit of the heterotic string. However, as we will show
\cite{us-next}, the form of the ten-dimensional 
effective theory of the strongly coupled heterotic string is actually
identical to that of the one-loop weakly coupled limit. Thus, since we
expect quantization conditions to be independent of coupling, 
equation (\ref{eq:G}) should correspond to the flux quantization condition of
the weakly coupled heterotic string. Comparing against the flux quantization
condition derived in ref.~\cite{rw}, we see that this is indeed the case.
We conclude, therefore, that expression (\ref{eq:B}) is the correct flux
quantization law for any four-cycles in Ho\v rava--Witten theory.
This argument could be made stronger by considering the membrane
action. For open membranes which end on the orbifold planes, one expects
addition fields in the world-volume theory which live only on the
boundary of the membrane. It is the coupling of these fields to the
gauge fields on the orbifold planes that one expects to lead to the
additional boundary terms in the quantization condition.

\vspace{0.4cm}

We are now ready to apply the quantization rule which we have just presented
to gaugino condensation. To do that, we consider a cycle $C_4=S^1\times C_3$
where $C_3$ is the three--cycle in the Calabi--Yau space which is nonzero
upon integration over $\e_{abc}$ and zero upon integration over the
harmonic $(2,1)$--forms $\o^{(i)}_{ab\bar{c}}$. Then, the components of
the four--form $G$ which contribute to the integral are $G_{abc11}$.
Recalling the definition of $\tilde{G}$ in terms of $G$, eq.~\refs{Gtdef},
and the form of $\tilde{G}$, eqs.~\refs{Gtilde}, \refs{Gc} we have for
these components~\footnote{Note that the only other part of $\tilde{G}$
with a $(abc11)$--component is $G^{(B)}_{abc11}$ given in eq.~\refs{GB}.
Since we have assumed that the matter fields $C^p$ are small fluctuations
it does not contribute to the quantization. In fact, even if it was
included, we would find it was cancelled by a corresponding piece in
the third term in the quantization condition, since both arise as the
dimensional reduction of Chern-Simons terms.}
\be
 G_{abc11} = -\frac{\sqrt{2}}{16\p}\left(\frac{\k}{4\p}\right)^{2/3}\left[
             \frac{1}{2\p\r}(\L^{(1)}+\L^{(2)}+\l )-(\L^{(1)}\d (x^{11})
             +\L^{(2)}\d (x^{11}-\p\r ))\right]\e_{abc}\; . 
\ee
Furthermore, we note that for our background and the specific cycle we have
chosen, the second and third integrals in the quantization rule do not
contribute. We therefore have
\be
  \zeta \int_{C_4}G=2\p n\; .
\ee
Inserting the above expression for $G_{abc11}$, the condensate--dependent
pieces cancel and we find a condition on $\l$ which reads
\be
 \l = \frac{32\p^2}{c\sqrt{V}}n \label{lquant}
\ee
where $c$ is an order one quantity defined by the relation
$\int_{C_3}\e_{abc}dx^a\wedge dx^b\wedge dx^c = c\sqrt{V}$ and $n$ is
an integer as before. 

Let us now discuss the implications of this result for the
potential~\refs{pot}. From now on we will concentrate on the ``physical''
case of gaugino condensation in the hidden sector; that is, we will
assume that $\L^{(1)}=0$ and $\L^{(2)}\neq 0$. For a nonzero integer $n$
we have a minimum of the potential~\refs{pot} at $\L^{(2)}=\l$, which
leads to soft supersymmetry breaking terms of order
$m_{\rm soft}\sim G_N\l\sim G_Nn/\sqrt{V}$. For a realistic value
$V^{-1/6}\sim 10^{16} $GeV, this implies soft breaking terms far too
large to be compatible with low energy supersymmetry. Therefore, the only
possible choice for the integer $n$ is $n=0$, which implies
\be
 \l = 0\; .
\ee
We have, therefore, seen that the arbitrary parameter in the low energy
potential can be fixed by the M--theory quantization rule in a way which
is very similar to the weakly coupled case~\cite{rw}.

To write the potential~\refs{pot} in an explicitly moduli dependent
form, we note that the condensates $\L^{(i)}$ correspond to a nonstandard
normalization of the four--dimensional gaugino kinetic terms. Taking this
into account, we have~\cite{drsw}
\be
 \L^{(2)}\sim\frac{1}{\sqrt{V}}\exp\left[-\frac{6\p}
             {b_0\a_{\rm GUT}}(S-\b T)\right]
\ee
and we get for the final potential
\be
 S^{(\c )} \sim \frac{\k^2}{\r V^2}\int_{M_4}e^{-12a-3c}\left|
                \exp\left[-\frac{6\p}{b_0\a_{\rm GUT}}(S-\b T)\right]
                \right|^2\; .
\ee
Up to power law corrections, this leads to the superpotential
\be
 W^{(\c )} = h\exp\left[-\frac{6\p}{b_0\a_{\rm GUT}}(S-\b T)\right]
\ee
where $h$ is a constant of order $\k/\r^{1/2}V$. This result exactly
coincides with the one in the weakly coupled case~\cite{drsw} if one takes
into account that the expression~\refs{couplings} for the gauge kinetic
function also holds in the weakly coupled region. This contradicts
claims~\cite{lt} that the potential is significantly more complicated
on the M--theory side. The crucial difference to the weakly coupled case
is not the form of the potential, but the size of the parameter $\b$
which is small for weak coupling but potentially large for strong
coupling. It is also interesting to note that, for fixed $S$ (fixed
Calabi--Yau radius), $T$ is driven toward small values; that is,
toward the weak coupling region~\cite{bd}.

\section{Soft terms}

As a final application, we would like to calculate the generic pattern of
soft terms that arises in the strongly coupled heterotic string. A similar
analysis has been carried out in ref.~\cite{hp}.

Let us first discuss the possible goals and limitations of such a computation.
As we have seen in the previous section, our simple gaugino condensate
potential shows a runaway behaviour, as in the weakly coupled case,
and does not provide us with a minimum for the moduli fields. Therefore,
unless we go to more complicated models, like, for example, multi--gaugino
condensation~\cite{multi}, the specific structure of supersymmetry breaking
is not determined. In this paper, we will not attempt to explicitly construct
such realistic models, but simple assume that supersymmetry breaking can be
achieved in more complicated cases. Though we do not gain any information
about the specific breaking pattern, we can still parameterize the effect
of supersymmetry breaking by the auxiliary components $F^S$, $F^T$ of
the moduli superfields $S$ and $T$. The interesting new feature
which motivates an analysis of soft terms, even without detailed knowledge
of the supersymmetry breaking mechanism, is the appearance of the correction
terms proportional to $\b$ in the K\"ahler potential and the gauge kinetic
functions~\refs{KWf}. We stress again that these terms
are present in the weakly coupled case as well. The important difference
is the magnitude of these terms in the weakly and strongly coupled case.
Let us adopt a normalization where the moduli field $S$, $T$ take values
of order one, so that the parameter $\b$ describes the order of magnitude
of the correction. Then $\b = O(\k^{2/3}\r /V^{2/3})$ is small in the
weakly coupled regime, but it can be sizeable and even of the order one or
larger in the strongly coupled regime. In fact, for the ``phenomenological''
values of $\r$ and $V$ determined from the Newton constant and the
grand unification coupling constant via eq.~\refs{couplings}, it is of
order one. The precise value depends somewhat on the specific Calabi--Yau
space chosen and can be computed from the equations presented in
ref.~\cite{low}. It should be kept in mind that the low energy theory
specified by eq.~\refs{KWf} is constructed as an expansion in $\b$ (up to
the first order) and, consequently, breaks down if $\b$ is too large.
We will, therefore, assume that $\b$ is still small enough so that the
expressions in~\refs{KWf} are sensible. Clearly, given these remarks, a
computation of soft terms only makes sense up to terms linear in $\b$. 

Having said this, the main goal of this section is to determine
the $O(\b )$ corrections to the soft terms which result from the
$\b$ dependent corrections to the K\"ahler potential and the
gauge kinetic function. In particular, it is interesting to see to what
extent the structure of soft terms, which is known to be somewhat special
for weakly coupled, large radius Calabi--Yau compactifications, is enriched
by those corrections.

\vspace{0.4cm}

The general structure of soft terms for hidden sector supersymmetry breaking
has been computed in ref.~\cite{sw,gm}. Here, we will use the convenient
K\"ahler--covariant approach and the notation of ref.~\cite{kl}.
We should point out that the transmission of supersymmetry breaking from
the hidden to the observable sector is completely governed by
four--dimensional supergravity as long as the theory is probed with momenta
far below the orbifold and the Calabi--Yau scale. In particular, at
such low momenta, the orbifold hyperplanes cannot be resolved and there
is no notion of being on the ``observable hyperplane'' receiving 
information about the breaking from the ``hidden plane''. Consequently,
the suppression of the breaking scale from the hidden to the observable
sector is not governed by the orbifold scale, but by the low energy
Planck scale. The auxiliary fields $F^S$ and $F^T$ are, therefore, of the
order $G_N\L^{(2)}$. For a realistic scenario we should therefore have
$(\L^{(2)})^{1/3}\sim 10^{13}$ to $10^{14}$ GeV resulting in
in a gravitino mass $m_{3/2}\sim F^T\sim 10^3$ GeV.

Let us now rewrite the model in a form adequate as a starting point for
the computation of soft masses. We introduce a four--dimensional
$\k$--parameter defined by $\k_P^2=8\p G_N=\k^2/V\p\r$. In order to
have matter fields $C$ of mass dimension one, we apply the rescaling
$C\rightarrow\k_PC$. Then the K\"ahler potential~\refs{KWf}, expanded
up to second order in $C$, can be written as
\be
 K = \k_P^{-2}\hat{K}(S,T,\bar{S},\bar{T})+Z_{\bar{p}q}(S,T,\bar{S},\bar{T})\,
     \bar{C}^{\bar{p}}C^q
\ee
with
\be
 \hat{K} = -\ln (S+\bar{S})-3\ln (T+\bar{T})\; ,\qquad
 Z_{\bar{p}q} = \left(\frac{3}{T+\bar{T}}+\frac{\b}{S+\bar{S}}\right)
                \d_{\bar{p}q}\; .
\ee
The rescaled superpotential is given by
\be
 W=\frac{1}{3}\tilde{Y}d_{pqr}C^pC^qC^r
\ee
with a coupling $\tilde{Y}$ of order one. Finally we need the gauge kinetic
function of the observable $E_6$ gauge group
\be
 f^{(1)} = S+\b T\; .
\ee
Inserting these quantities into the formulae presented in ref.~\cite{kl},
we arrive at the following tree level soft terms~\footnote{It is assumed
that the cosmological constant in the hidden sector vanishes.}
\bea
 m_{3/2}^2 &=& \frac{|F^S|^2}{3(S+\bar{S})^2}+\frac{|F^T|^2}{(T+\bar{T})^2}
               \nn \\
 m_{1/2} &=& \frac{F^S}{2(S+\bar{S})}+\b\left[\frac{F^T}{2(S+\bar{S})}-
             \frac{T+\bar{T}}{(S+\bar{S})^2}F^S\right] \nn \\
 m_0^2 &=& \frac{|F^S|^2}{(S+\bar{S})^2(T+\bar{T})}+\b\left[
           -\frac{5|F^S|^2}{3(S+\bar{S})^3}+\frac{2\mbox{Re}
           (F^S\bar{F}^{\bar{T}})}{(S+\bar{S})^2(T+\bar{T})}\right] \\
 A &=& -\left[\frac{F^S}{S+\bar{S}}+\b\left(\frac{F^T}{S+\bar{S}}-
       \frac{T+\bar{T}}{(S+\bar{S})^2}F^S\right)\right]Y\nn
\eea
where $Y=e^{\hat{K}/2}\tilde{Y}$. Here, $m_{3/2}$, $m_{1/2}$, $m_0$ and $A$
are the gravitino mass, the gaugino mass, the scalar mass and the trilinear
coupling. (Note that the scalar masses are not quite correctly normalized
in these expressions, since we have not rescaled the K\"ahler metric out
of the corresponding kinetic terms.) Let us first consider the above
expressions in the limit $\b\rightarrow 0$. In this case, we recover the
expression for the soft masses one obtains in a weakly coupled, large radius
Calabi--Yau compactification~\cite{bim}. In this limit, $m_{1/2}$, $m_0$
and $A$ depend on the auxiliary component $F^S$ only. This specific structure
results from the form of the K\"ahler potential and (as far as the
trilinear coupling is concerned) from the fact that the Yukawa couplings are
constant in the large radius limit. Experience with gaugino condensation
in the weakly coupled case shows that supersymmetry is broken frequently
in the $F^T$ direction only,  in which case all soft couplings except
$m_{3/2}$ would be small.

Let us now discuss the effect of the additional terms proportional to $\b$.
A priori, one could expect the $F^T$ degeneracy of all couplings to be
lifted by those corrections. While this is true for the gaugino masses
$m_{1/2}$, as has been pointed out in ref.~\cite{hp}, and the trilinear
coupling $A$, as we emphasize in this paper, the scalar
soft masses still receive no contribution proportional to $|F^T|^2$.
Consequently, for supersymmetry breaking in the $F^T$--direction, the
scalar soft masses remain light in the M--theory regime while $m_{1/2}$
and $A$ are of the same order as $m_{3/2}$. We note that
the soft couplings also receive various new contributions proportional
to $F^S$ which will be relevant for the precise value of the couplings.
They will, however, not cause a change in the order of magnitude in
going from the weakly to the strongly coupled regime.

\vspace{0.4cm}

As this manuscript was prepared, ref.~\cite{ckm} appeared. This paper
discusses some of the issues presented in this work, particularly the
phenomenological results of section 5.

\vspace{0.4cm}

{\bf Acknowledgments} A.~L.~is supported in part by a fellowship from
Deutsche Forschungsgemeinschaft (DFG). A.~L.~and B.~A.~O.~are
supported in part by DOE under contract No. DE-AC02-76-ER-03071.
D.~W.~is supported in part by DOE under contract No. DE-FG02-91ER40671.

\end{document}